\documentclass[11pt,a4paper]{article} 
\usepackage{style}

\usepackage{amsmath,amssymb,epsfig,bm,amsfonts,bbm,mathrsfs}

\newcommand{\ket}[1]{| #1 \rangle}

\newcommand{\braopket}[3]{\langle #1 | #2 | #3 \rangle}
\newcommand{\mat}[1]{\begin{pmatrix} #1 \end{pmatrix}}

\relax
\title{Neutrino flavor-mass uncertainty relations and an entanglement-assisted determination of the PMNS matrix}

\author{Stefan Floerchinger}
\emailAdd{stefan.floerchinger@thphys.uni-heidelberg.de}
\affiliation{Institut f\"{u}r Theoretische Physik, Universit\"{a}t Heidelberg, Philosophenweg 16, 69120 Heidelberg, Germany}
\author{and Jan-Markus Schwindt}
\emailAdd{jan.schwindt@gmail.com}
\abstract{As a result of a non-trivial mixing matrix, neutrinos cannot be simultaneously in a flavor and mass eigenstate. We formulate and discuss information entropic relations that quantify the associated quantum uncertainty. We also formulate a protocol to determine the Pontecorvo-Maki-Nakagawa-Sakata (PMNS) matrix from quantum manipulations and measurements on an entangled lepton-neutrino pair. The entangled state features neutrino oscillations in a conditional probability involving measurements on the lepton and the neutrino. They can be switched off by choosing a specific observable on the lepton side which is determined by the PMNS matrix. The parameters of the latter, including the CP-violating phase $\delta$, can be obtained by guessing them and improving the guess by minimizing the remaining oscillations.}
\begin{document}
\maketitle
\section{Introduction}

Heisenberg's uncertainty relation \cite{aHeisenberg:1927zz} is one of the most famous results of quantum mechanics.
Applied to position and momentum, it gives a simple and state-independent lower bound on the product of the uncertainties of these two observables \cite{Kennard}, and thereby defines a central feature of wave-particle duality. Generalized to arbitrary pairs of observables \cite{Robertson:1929zz,Schroedinger}, Heisenberg's uncertainty relation is still defined, but is state-dependent in most cases, and doesn't give constraints as strong and useful as in the position/momentum case, unless the chosen observables are ``complementary" in the sense that the commutator of the operators representing the observables is a multiple of the identity.   

For observables with a finite spectrum (like spin, for example), a different type of uncertainty can be defined in terms of information entropy, which is often more useful in these cases. Similar to Heisenberg's uncertainty relation, inequalities can be formulated for given observables that constrain their associated information-entropic uncertainties \cite{Deutsch,Kraus,Maassen:1988ryr,Frank:2012hz}. These entropic uncertainty relations have been successfully applied to a number of problems in the field of quantum information and computation \cite{Coles:2017,Wilde}. Extensions to observables with continuous spectrum (e.g. position and momentum) are also possible.

In particle physics, this concept is not yet very well-known, but may have some useful applications. Neutrino physics (see refs.\ \cite{King:2014nza,Verma:2015tpa,Tanabashi:2018oca} for an overview) is a particularly interesting area in this regard: The Hilbert space $\mathcal{H}$ of a neutrino has a three-dimensional factor 
$\mathcal{H}_f$ which accounts for its state with respect to flavor as well as mass. In other words, flavor and mass are two observables that ``live" in $\mathcal{H}_f$. From the existence of neutrino oscillations, we know that these
observables do not have simultaneous eigenstates. A flavor eigenstate will imply some uncertainty in
mass, and vice versa. This is a relatively simple case where entropic uncertainty relations can be applied.

Our intention in this regard is here twofold. First, we want to bring the notion of entropic uncertainty
to the attention of the particle physics community, applying it to the simple example of 
the neutrinos' flavor/mass space $\mathcal{H}_f$. Second, we study possible applications in neutrino physics where this technique might turn out useful.

Another very important concept in quantum information theory is entanglement, of course. For neutrino physics it can play a rather interesting role as well. We illustrate this by discussing how one may determine the Pontecorvo-Maki-Nakagawa-Sakata (PMNS) matrix from quantum manipulations and measurements on an entangled lepton-neutrino state as it could result from an electroweak decay. We will also discuss extensions of the entropic uncertainty relations to such situations with ``side information'' from entanglement, also known as ``quantum memory''. (A different type of entanglement in the context of neutrino oscillations has been discussed previously in refs.\ \cite{Blasone:2007vw,Blasone:2007wp,Blasone:2014jea,Alok:2014gya,Banerjee:2015mha}.)

The paper is organized as follows: In section 2, we introduce the general concept of entropic uncertainty,
based on information theory, and compare it to the type of uncertainties that were used by Heisenberg and followers, based
on standard deviations. We also compare the resulting uncertainty relations. In section 3, we apply both relations
to the mass uncertainty of a neutrino produced in a pion decay, and show that the entropic version is
more useful. In section 4, possible applications in neutrino physics are discussed, in particular the areas of quantum computation 
and cosmology. In section 5, we extend the
discussion to entangled states of leptons and neutrinos, and show how measurements on the lepton can enhance
our knowledge about the neutrino. In particular, we discuss how entanglement can be used for a determination of the PMNS matrix via quantum manipulations and measurements on a lepton-neutrino pair, and we provide an extension of entropic uncertainty relations to situations with ``quantum memory''. In section 6, we present our conclusions.  
\section{Heisenberg vs. entropic uncertainty relations}

Heisenberg's uncertainty relation \cite{aHeisenberg:1927zz} for position and momentum of a non-relativistic particle described by quantum mechanics is a direct consequence of quantization. More specifically, from the operator commutation relation
\begin{equation}
[\hat x_i, \hat p_j] = i \hbar \delta_{ij},
\end{equation}
one derives for the standard deviations $\Delta x_i = \langle (x_i - \langle x_i\rangle)^2 \rangle^{1/2}$ (and similar for momentum) the inequality \cite{Kennard} 
\begin{equation}
\Delta x_i \Delta p_j \geq \frac{1}{2} \hbar \delta_{ij}.
\end{equation}

This relation is state-independent, since the commutator is a multiple of the identity operator.
More generally, for two observables $\hat z$, $\hat x$ and quantum state $\ket{\psi}$ one has a state-dependent inequality for their 
standard deviations $\sigma_{z,\psi}$ and $\sigma_{x,\psi}$ in terms of the expectation values of their commutator 
and anticommutator \cite{Schroedinger}
\begin{equation}
 \sigma_{z,\psi} \sigma_{x,\psi} \geq \frac{1}{2}
 \sqrt{\braopket{\psi}{\{\hat z - \langle \hat z \rangle_\psi, \hat x - \langle \hat x \rangle_\psi\}}{\psi}^2
  + |\braopket{\psi}{[\hat z, \hat x]}{\psi}|^2}.
 \label{eq:FullHeisenberg}
\end{equation}
The term involving the anticommutator is generically of the same order of magnitude as the commutator term, but is more difficult 
to compute (expectation values need to be determined). In the canonical case of position and momentum, it vanishes.
The term is usually dropped to allow for a simpler relation, at the cost of getting a weaker bound.
This simplified inequality \cite{Robertson:1929zz},
\begin{equation}
\sigma_{z,\psi} \sigma_{x,\psi} \geq \frac{1}{2} | \braopket{\psi}{[\hat z, \hat x]}{\psi} |,
\label{eq:Robertson}
\end{equation}
is known as the general Heisenberg, or Robertson, uncertainty relation.

An alternative way to formulate uncertainty relations uses the language of information theory and entropy. Note that Heisenberg uncertainty
relations are valid for operators with discrete or continuous spectrum, while the following relations can be expressed most directly for 
operators with discrete spectrum, although generalizations to the continuous case are possible. 
To formulate entropic uncertainty relations in general terms, we consider again two observables $\hat z$ and $\hat x$ with normalized sets of eigenfunctions $| z_j \rangle$ and $|x_k \rangle$, respectively. For a given density matrix $\rho$, the probabilities for different outcomes in measurements of $\hat z$ and $\hat x$ are given by $p^{(z)}_j = \langle z_j | \rho | z_j \rangle$ and $p^{(x)}_k = \langle x_k | \rho | x_k \rangle$, respectively. If the two observables $\hat z$ and $\hat x$ are not compatible (i.e. do not commute), they can not both be sharp at the same time, meaning that for any density matrix $\rho$ the uncertainty with respect to $\hat z$ and the uncertainty with respect to $\hat x$ cannot both vanish. This will be quantified in terms of entropic uncertainty relations.

An interesting feature of these relations is that they are formulated solely in terms of probability distributions $p^{(z)}_j$ and $p^{(x)}_k$, respectively. First, the information content associated with the outcome $j$ for the observable $x$ is given by
\begin{equation}
i^{(z)}_j = - \ln p^{(z)}_j.
\end{equation}
This is the amount of information one learns from a measurement of the variable $\hat z$ with outcome $z_j$. The Shannon or information entropy associated with the probability distribution for the measurement of $\hat z$ is given by the expectation value of the information content,
\begin{equation}
H^{(z)} = \langle i^{(z)} \rangle = - \sum_j p^{(z)}_j \ln p^{(z)}_j.
\end{equation}
This is obviously a quantitative measure for the uncertainty associated to the variable $\hat z$: if the probability distribution $p^{(z)}_j$ is sharp such that $p^{(z)}_j=1$ for one value of $j$, the information entropy $H^{(z)}$ vanishes, whereas $H^{(z)}$ is maximal for a flat probability distribution.
In a fully analogous way, the information entropy associated with a measurement of $x$ is given by
\begin{equation}
H^{(x)} = \langle i^{(x)} \rangle = - \sum_k p^{(x)}_k \ln p^{(x)}_k.
\end{equation}

To formulate entropic uncertainty relations, one also needs the maximal overlap between the basis functions, defined by
\begin{equation}
c = \sup_{j,k} | \langle z_j | x_k \rangle |.
\end{equation}
Note that the maximal overlap is a real number in the range $0< c \leq 1$ and that it adopts its maximal value $c=1$ if the basis sets have a common eigenvector (possibly up to an irrelevant phase factor) such that $|\langle z_j | x_k \rangle| = 1$ for some value of $j$ and $k$.

A final ingredient for the entropic uncertainty relation is the von Neumann entropy associated with the full density operator $\rho$,
\begin{equation}
S = - \text{Tr} \{ \rho \ln(\rho) \}.
\end{equation}
For pure states, the von Neumann entropy $S$ vanishes but is positive for mixed states.

With these ingredients, one can write an entropic uncertainty relation \cite{Deutsch,Kraus,Maassen:1988ryr} for the measurements of $z$ and $x$ as \cite{Frank:2012hz}
\begin{equation}
H^{(z)} + H^{(x)} \geq 2 \ln(1/c) + S.
\label{eq:FrankLieb2002}
\end{equation}
Indeed, the left hand side measures the combined uncertainty of the probability distributions for the measurements of $\hat z$ and $\hat x$, respectively. The first term on the right hand side is state-independent and measures the ``incompatibility'' of the observables $\hat z$ and $\hat x$, respectively. In particular, it vanishes if the basis sets $| z_j \rangle$ and $|x_k \rangle$ share a state and it is large if the overlap between them is small. Finally, the von Neumann entropy $S$ on the right hand side of \eqref{eq:FrankLieb2002} quantifies the additional uncertainty that arises for a mixed state.

\section{Uncertainties in neutrino production from weak decays}

Let us now discuss the application of the general uncertainty relations to neutrino flavor and mass quantum numbers. We consider neutrinos with flavor eigenstates $| \nu_\alpha \rangle$, labeled by the flavor index $\alpha \in \{e, \mu, \tau\}$, and mass eigenstates $|\nu_j\rangle$ with the mass index $j\in\{1,2,3\}$. The relation between both is given in terms of the Pontecorvo-Maki-Nakagawa-Sakata (PMNS) matrix \cite{Pontecorvo:1957qd,Maki:1962mu}, \footnote{We assume here that the PMNS matrix is unitary. This is not necessarily the case in the presence of right-handed, sterile neutrinos, see e.\ g.\ ref.\ \cite{King:2014nza}.}
\begin{equation}
| \nu_\alpha \rangle = U^*_{\alpha j} |\nu_j \rangle.
\end{equation}

As a concrete example, we consider a neutrino produced from the process $\pi^+ \to \mu^+ + \nu_\mu$.
The resulting neutrino state is a superposition of mass eigenstates, which was experimentally found by the observation
of neutrino flavor oscillations. The presence of these oscillations requires a number of kinematic uncertainties, as the neutrino
has to be simultaneously on different mass shells $E_i^2 - p_i^2 = m_i^2$. These kinematic uncertainties have been studied in detail
in ref.\ \cite{Akhmedov:2009rb}. Here, however, we will focus only on uncertainties in the three-dimensional mass/flavor space.  
We want to determine the neutrino's mass uncertainty in the state $\ket{\nu_\mu}$.
We will do so in three ways and compare the results: (i) directly from the quantum state, (ii) via
Heisenberg's uncertainty relation \eqref{eq:Robertson}, and (iii) via the Entropic uncertainty relation \eqref{eq:FrankLieb2002}.

\paragraph{(i) Directly from quantum state.} From neutrino oscillation experiments, we roughly know the absolute values of the matrix elements $U_{\alpha j}$, but not the phases.
We also roughly know the differences, 
\begin{equation}
 \Delta m_{ij}^2 = m_i^2 - m_j^2 = (m_i - m_j)(m_i + m_j),
\end{equation}
but not the absolute masses. The definition of the standard deviation gives a mass uncertainty of
\begin{equation}
 \sigma_{m,\mu} = \sqrt{p_1 [ p_2(m_1-m_2) + p_3(m_1-m_3) ]^2 + \text{cycl. perm.}},
\end{equation}
where the $p_j$ are the probabilities to find $m_j$ in the muon flavor eigenstate $| \nu_\mu \rangle$. With
\begin{equation}
 \ket{\nu_\mu} = \sum_j U_{\mu j}^* \ket{\nu_j},
\end{equation}
one has $p_j = |U_{\mu j}|^2$. The phases are not needed, and the known absolute values result in $p_1 \approx 1/6$,
$p_2 \approx 1/3$, $p_3 \approx 1/2$, approximately. However, the mass differences $m_i - m_j$ are not known, only differences of the squared
masses. Therefore, we cannot calculate the mass uncertainty based on the given information.
In order to proceed, we can make an additional assumption on the hierarchy of the neutrino masses: we assume that it is similar to the
hierarchy of the corresponding leptons, i.e. the lightest neutrino is much lighter than the other two, and the second still an order
of magnitude lighter than the third. This is consistent with the observed values
\begin{equation}
 \Delta m_{21}^2 \approx 7.5 \times 10^{-5} \text{ eV}^2 , \qquad 
 \Delta m_{32}^2 \approx \Delta m_{31}^2 \approx 2.5 \times 10^{-3} \text{ eV}^2.
\end{equation}
In this case, we can set $m_1 \approx 0$, $m_2 \approx 10 \text{ meV}$, $m_3 \approx 50 \text{ meV}$, which gives an uncertainty
$\sigma_{m,\mu} \approx 22 \text{ meV}$.

The information entropy associated with the mass measurement, 
on the other hand, can be directly read off from the probabilities, without any extra assumptions on the
masses,
\begin{equation}
  H^\text{Mass} = -\sum p_i \ln p_i \approx 1.01.
\end{equation} 

\paragraph{(ii) Heisenberg uncertainty relation.} In principle, a lower bound on the neutrino mass uncertainty can be determined from the Heisenberg uncertainty relation via
\begin{equation}
 \sigma_{m,\mu} \geq \frac{|\braopket{\psi}{[M,F]}{\psi}|}{\sigma_{f,\mu}}.
 \label{eq:SigmaMFromHeisenberg}
\end{equation}
However, this calculation faces a number of problems, and eventually gives only a bad approximation.

At first we need to define mass and flavor operators. In the mass basis, the mass operator $M$ is obviously given by
\begin{equation}
 M_{ij} = \text{diag}(m_1,m_2,m_3).
\end{equation}
The flavor operator $F$ is diagonal in the flavor basis, but what eigenvalues shall we assign it? A natural choice (considering
flavor as an approximate symmetry group) seems
\begin{equation}
 F_{\alpha\beta} = \text{diag}(-1,0,1).
\end{equation}
Note that a linear shift $F \to aF+b$, will not change the result of \eqref{eq:SigmaMFromHeisenberg}, but a non-linear change of the
eigenvalues will. This can be considered a first problem of this approach: the bound to be derived
depends not only on the quantum state, but also on the eigenvalues assigned to an observable (flavor, in this case) which is not
{\it per se} numeric.

Then, to compute the commutator, we need to have $M$ in the flavor basis, which requires complete knowledge of $U_{\alpha j}$,
including all phases (which are unknown, as mentioned above).
This is the second problem. In order to move forward, we take into account that the experimental findings for 
$U$ are roughly consistent with $U$ being a tri-bimaximal mixing matrix (assuming vanishing CP violating phase $\delta=0$) \cite{Harrison:2002}
\begin{equation}
 U_{\alpha j} \approx \mat{\sqrt{\frac{2}{3}} & \sqrt{\frac{1}{3}} & 0 \\
                           -\sqrt{\frac{1}{6}} & \sqrt{\frac{1}{3}} & \sqrt{\frac{1}{2}} \\
													 \sqrt{\frac{1}{6}} & -\sqrt{\frac{1}{3}} & \sqrt{\frac{1}{2}} },
\end{equation}
and for now we simply assume that $U$ has this form. This allows us to determine 
\begin{equation}
 M_{\alpha\beta} = U_{\alpha i}M_{ij}U^\dagger_{j \beta},
\end{equation}
and from there
\begin{equation}
 [M,F]_{\alpha\beta} = \frac{1}{6} \mat{ 0 & -2m_1+2m_2 & 4m_1-4m_2 \\
                                         2m_1-2m_2 & 0 & -m_1-2m_2+3m_3 \\
																				 -4m_1+4m_2 & m_1+2m_2-3m_3 & 0 }.
\end{equation}
Now, for a flavor eigenstate $\ket{\nu_\mu}$, the flavor uncertainty $\sigma_{f,\mu}$ vanishes as well as the
value $\braopket{\nu_\mu}{[M,F]}{\nu_\mu}$, so \eqref{eq:SigmaMFromHeisenberg} amounts to a division 0/0. This is
our third problem. To solve it, we remember that $\pi^+$ also has a decay channel $\pi^+ \to e^+ + \nu_e$,
which is however strongly suppressed by helicity and phase space constraints. We may therefore consider our
neutrino state as
\begin{equation}
 \ket{\nu} = \left[1-{\cal O}\left(|\xi|^2\right)\right]\ket{\nu_\mu} + \xi\ket{\nu_e}
\end{equation}
with a tiny complex value $\xi$. Now we find
\begin{equation}
 \sigma_{f,\nu} = -|\xi|, \qquad \braopket{\nu}{[M,F]}{\nu} = \frac{2}{3}(m_1-m_2) \text{Im}\,\xi
\end{equation}
and \eqref{eq:SigmaMFromHeisenberg} gives
\begin{equation}
 \sigma_{m,\nu} \geq \frac{2}{3}(m_2-m_1) \frac{\text{Im}\,\xi}{|\xi|}.
\end{equation}
It seems odd that our bound depends on the phase of $\xi$, and also that the mass $m_3$ does not show up. This is actually
a consequence of dropping the anticommutator term (our fourth problem) from \eqref{eq:FullHeisenberg}. Including it gives a
somewhat better bound, but also a more complicated calculation. 

Altogether, it turns out that Heisenberg's uncertainty relation is quite useless for our case, for a number of reasons. It is just 
much simpler and better defined to read off the uncertainty from the state directly.

\paragraph{(iii) Entropic uncertainty relation.} Finally, we evaluate the entropic uncertainty relation \eqref{eq:FrankLieb2002} for the neutrino flavor/mass mixing.
The maximal overlap between flavor and mass eigenstates is given by the maximal matrix component of the PMNS matrix,
\begin{equation}
c_\text{PMNS} = \sup_{\alpha,j} |\langle \nu_\alpha | \nu_j \rangle| = \sup_{\alpha,j} |U_{\alpha j}^*|.
\end{equation}
A one-particle state described by the one-particle density matrix $\rho$ is then subject to the flavor-mass uncertainty relation
\begin{equation}
H^\text{Flavor} + H^\text{Mass} = 2 \ln(1/c_\text{PMNS}) + S.
\end{equation}
In particular, from neutrino oscillation experiments, it turns out that the maximal PMNS matrix component is $c_\text{PMNS}=U_{e1} = 0.82 \pm 0.01$ \cite{Tanabashi:2018oca}. For a pure state with $S=0$ one infers the flavor-mass uncertainty relation
\begin{equation}
H^\text{Flavor} + H^\text{Mass} \geq 2 \ln(1/c_\text{PMNS}) = 0.39 \pm 0.02.
\end{equation}
We emphasize that this bound is independent of the specific state considered and a direct consequence of a non-trivial neutrino mixing matrix. 
For example, if a neutrino is prepared in a flavor eigenstate $|\nu_\mu\rangle$ (as in our $\pi^+$ decay), 
the corresponding information entropy vanishes, $H^\text{Flavor}=0$. The information entropy associated with a measurement of mass is then bounded from below by $H^\text{Mass} \geq 0.39 \pm 0.02$. The fact that flavor and mass measurements are incompatible expresses itself in $2 \ln(1/c_\text{PMNS})$ being larger than zero. On the other side, the numerical value is still relatively small against the maximal possible value $\ln(1/c)=\ln(3)\approx 1.10$ which would correspond to maximal incompatibility between flavor and mass (in the sense that mass is completely uncertain for known flavor and vice versa). 

For the muon neutrino state $|\nu_\mu\rangle$ we found $H^\text{Mass} \approx 1.01$, which is quite close to the maximal value and quite far away from the bound
given by the uncertainty relation. The entropic uncertainty relation \eqref{eq:FrankLieb2002}, in contrast to the generic Heisenberg
uncertainty relation, does not depend on the state. This is a great simplification, but it comes at the cost of getting
rather weak bounds for certain states, as $c_\text{PMNS}$ is only based on the flavor state 
(which happens to be the electron neutrino) which maximally overlaps with a mass state.   

The calculation above shows how much easier the information entropy can be handled, in the case of discrete observables, as compared
to uncertainties based on standard deviations. However, the question remains in what cases and for what purposes
information-theoretic concepts like entropic uncertainty can be useful in the context of neutrino physics. 
\section{Applications}
Entropic uncertainty can be a useful concept in situations where we one has to deal with
statistical distributions of quantum states, when entropy- or information-related quantities are
of specific interest. In the case of individual states,
information entropy can be read off from the state directly, as we have seen in 
the case of the muon in the previous section. When an ensemble of states is given, entropic uncertainty relations
can be used. On the other hand, in the case where one is interested only in expectation values
and possibly standard deviations of physical quantities, information entropy is not of primary interest,
and can be at best of indirect use. Keeping this in mind, we identify the following areas of neutrino
physics where information-theoretic concepts can be useful:
\begin{enumerate}
 \item {\bf Quantum communication and computation:} This is, of course, the standard arena for quantum information.
  From a theory perspective, neutrinos are good candidates for quantum computation
  applications, since they are, thanks to their tiny interaction cross sections, almost free from
	decoherence effects. Information can, in principle, be encoded in the three-dimensional flavor/mass space.
	Several operations are possible to manipulate the corresponding states.
	With the observed flavor oscillations, we have a natural way of rotating a neutrino in flavor
	space (although not completely and not along arbitrary axes). With induced beta decay and its inverse,
	$\nu + n \longleftrightarrow \ell^- + p^+$ and $\ell^+ + n \longleftrightarrow \bar{\nu} + p^+$, 
	neutrinos can be transformed into charged leptons and vice versa, preserving the flavor state.
	Charged leptons can be kept in a coherent superposition of flavors, and using magnetic fields,
	flavor components can be separated and manipulated individually. The different mass components of 
	a neutrino correspond to wave packets with slightly different velocities, which will separate after
	certain travel distances (for the ultra-relativistic neutrinos used in nowadays experiments, however,
	these distances are very large), which can be used, in principle, to measure which mass state a neutrino is in, 
	or to manipulate different mass components separately.
	As for entangled states, pion decay $\pi^\pm \longrightarrow \mu^\pm + \nu_\mu$ serves as a source of
	an entangled muon/neutrino pair, while weak boson decay $W^\pm \longrightarrow \ell^\pm + \nu$
	produces entangled mixed-flavor pairs.
	
	These possibilities show that there is, in principle, a rich set of instruments available to manipulate or read quantum
	information in flavor space.
	On the other hand, the low interaction rates of neutrinos are also a disadvantage: either huge numbers
	of neutrinos or huge numbers of nuclei are needed to transfer, manipulate, or read a piece of information. It remains
	to be seen how applicable neutrino quantum computation is in practice. If it turns out to be viable, 
	entropic uncertainty relations will be
	of similar use as they are already in standard photonic/electronic/atomic quantum computation and communication.
	
 \item {\bf Cosmology:} Statistical processes and entropies play a big role in cosmology. In the {\it early universe}, 
  according to the Leptogenesis scenario \cite{Akhmedov:1998qx,Asaka:2005pn}, right-handed neutrinos and their oscillations 
	may have played a decisive role in producing the small matter/antimatter asymmetry (relative to the overall entropy). It is 
	conceivable that entropy and other information-theoretic concepts can help to gain an improved understanding of 
	such interesting theoretical scenarios in the future.
 
 	One second after the Big Bang, left-handed
	neutrinos decoupled from matter and now form a cosmic background similar to the CMB, rarely interacting with
	anything. In the {\it late universe}, that is, starting in the present era and even more so in the future, the
	different mass component of the cosmic neutrinos behave somewhat differently. While neutrinos cool down, they
	become non-relativistic at some point, but the time when this happens differs by mass component, 
	because heavier neutrinos have smaller
	velocities at the same temperature. Therefore, they start to participate in structure formation earlier, leading to
	a higher abundance of the heaviest component inside galaxies, and a lower one in empty space. The information entropy
	regarding cosmic neutrino masses is reduced. This will play a role in future experiments to measure the cosmic neutrino 
	background.
	
 \item {\bf Constraints on the PMNS matrix:} If we find a way to measure neutrino masses directly on a given neutrino state
  (e.g. by separating the wave packets of the various mass components), 
  we would be able to determine the information entropies regarding flavor and mass for a given ensemble of neutrinos, and can then 
	use entropic uncertainty relations to constrain parameters of $U_\text{PMNS}$, in particular the maximum entry $c$. 
	However, it is unlikely that 
	this will give better constraints than the established methods from neutrino oscillations, or than other, more direct methods that
	would also become available if neutrino masses can be separated.
	
	Instead, in the following section we will discuss a different approach how $U_\text{PMNS}$ can be constrained experimentally,
	including the CP-violating phase $\delta$, and how information-theoretic concepts can assist this approach.
\end{enumerate}

\section{Entanglement}
Entangled states are of particular interest in quantum information theory. If two objects A and B are entangled, we are able to 
acquire information about A by performing a measurement on B. Measurements on A will have smaller uncertainties when measurements on
B (the `side information') are taken into account.
For the neutrino case, consider an entangled state that could in principle result from some fully ``flavor blind'' decay process of the type $W^+\to l^+ + \nu$,
\begin{equation}
| \psi \rangle = \frac{1}{\sqrt{3}} \left( |e^+\rangle | \nu_e\rangle + |\mu^+\rangle | \nu_\mu\rangle + |\tau^+\rangle | \nu_\tau\rangle \right).
\label{eq:leptonneutrinoentangledstate}
\end{equation}
The neutrino is here fully entangled with a lepton $|l \rangle$, and one can therefore ``circumvent'' the uncertainty associated with
measurements on the neutrino by first measuring the associated observable on the lepton. Note that as it stands, 
\eqref{eq:leptonneutrinoentangledstate} is not an eigenstate of neutrino flavor and in fact, all three flavors have equal probability. However, if one takes side information into account which comes from a flavor measurement of the lepton, one can actually predict the outcome of the neutrino flavor measurement happening afterwards with certainty 
(at short distance, where neutrino oscillations can be neglected). This is a consequence of \eqref{eq:leptonneutrinoentangledstate} being fully entangled.

If it were possible to measure neutrino masses directly, the outcome of such a measurement could also be predicted from a quantum manipulation and measurement on the lepton, only. To see this, we write the state \eqref{eq:leptonneutrinoentangledstate} as
\begin{equation}
| \psi \rangle = \frac{1}{\sqrt{3}} \sum_{\alpha,\beta} \delta_{\alpha\beta} |l_\alpha^+\rangle  | \nu_\beta \rangle = \frac{1}{\sqrt{3}} \sum_{\alpha, \beta, j} U^*_{\alpha j} (U^T)_{j\beta}  |l_\alpha^+\rangle  | \nu_\beta \rangle = \frac{1}{\sqrt{3}} \sum_{\alpha,j} U^*_{\alpha j} |l_\alpha^+\rangle  | \nu_j \rangle = \frac{1}{\sqrt{3}} \sum_j | \tilde l_j^+ \rangle | \nu_j \rangle.
\label{eq:fullyentangledstate}
\end{equation}
This shows that the transformed lepton states $| \tilde l^+_j \rangle = \sum_\alpha U^*_{\alpha j} |l^+_\alpha \rangle$ are fully entangled with the mass eigenstates of the neutrinos $|\nu_j\rangle$. If one first applies a unitary operator corresponding to $U^{*T} = U^\dagger$ to the lepton (assuming for now that such a manipulation of the lepton is possible), and then measures its state, one can in fact predict the outcome of a neutrino mass measurement with certainty! 

These examples show that side information from an entangled state can reduce the uncertainty of a measurement or of two incompatible measurements. Note that this does not alter the fact that mass and flavor measurements are incompatible. After a mass measurement on
the neutrino, both the neutrino and the lepton will be in a state of uncertain flavor, and vice versa if flavor is measured.
It only means that the outcome of any measurement of the neutrino can be predicted if a corresponding measurement is first performed 
on the lepton (the `side information').

On the other hand, it is clear that a state where the lepton and neutrino parts factorize, $|\psi\rangle = | l \rangle |\nu\rangle$, is of an entirely different kind. Here a measurement in the lepton sector does not yield useful side information to predict the outcome of a measurement on the neutrino. Between these two extreme cases of full entanglement and the product state is a class of intermediate states with partial entanglement. 

In the following, we present an idea how to use the quantum entanglement between leptons and neutrinos to constrain the mixing matrix 
$U_\text{PMNS}$, in particular the so far undetermined CP-violating phase $\delta$ (subsection \ref{sec:Entanglementassistedmeasurement}). As a more general remark, we provide a generalization of the entropic uncertainty relation \eqref{eq:FrankLieb2002} when side information from entanglement (sometimes called quantum memory)
is taken into account (subsection \ref{sec:Entropicuncertaintyrelationwithquantummemory}).

\subsection{Entanglement assisted determination of the PMNS matrix}
\label{sec:Entanglementassistedmeasurement}

Consider again the fully entangled lepton-neutrino state in equation \eqref{eq:fullyentangledstate}. With the correct manipulation of the lepton consisting of a unitary operator $U^\dagger$ and a projective measurement, one can measure the states $|\tilde l_j^+ \rangle$. 
(We can speculate that the operation of $U^\dagger$ on the lepton is experimentally realized by a combination of magnetic fields 
and beam splitters.)
Let us assume that the result is given by the lepton state with index $j=k$. By entanglement, this implies a neutrino that is in a mass eigenstate 
$|\nu_k \rangle$. This can be experimentally confirmed even without the ability to measure neutrino masses: Neutrino oscillations are a 
consequence of the overlap of the various mass eigenstates, and so the absence of such an overlap results in an absence of neutrino
oscillations. For a neutrino in a mass eigenstate $|\nu_k \rangle$, the probability to find the neutrino in a flavor $\alpha$ is
given by $|U_{\alpha k}|^2$, independent of distance. 

In other words, for an ensemble of entangled states of the type \eqref{eq:fullyentangledstate}, one can perform projective measurements 
corresponding to the operators 
\begin{equation}
 P_j = | \tilde l_j^+\rangle \langle \tilde l_j ^+ |. 
\label{eq:projectionOperatorsLeptonMass}
\end{equation}
If one sorts the events according to the measurement outcome into three classes $j\in \{1,2,3\}$, for each class the corresponding neutrinos are in definite mass eigenstates and do not show any flavor oscillations. 
That is, we measure the leptons in 
the ``neutrino mass basis", and the neutrinos in the flavor basis (by having them interact with other leptons of determined flavor).
The correlations between the measurement results are then given by $|U_{\alpha k}|^2$, independent of distance.

In contrast, if the leptons are measured directly, i.\ e.\ with the projection operators 
\begin{equation}
 P_\alpha=| l^+_\alpha \rangle \langle l^+_\alpha |,
\label{eq:projectionOperatorsLeptonFlavor}
\end{equation}
and events then sorted with respect to the outcome $\alpha\in\{ e, \mu, \tau \}$, the neutrinos in each class are initially in flavor eigenstates and show the corresponding oscillations. 

The whole procedure described above only works if the correct matrix $U_\text{PMNS}$ is used to perform the quantum manipulations on the lepton side. One may now ask: Can this mechanism be used to constrain the form of $U_\text{PMNS}$? In the following we investigate this question in more detail. For concreteness we work with the standard parametrization
\begin{equation}
U_\text{PMNS} = V_\text{PMNS} \,  \Phi_\text{PMNS} = V_\text{PMNS} \begin{pmatrix} 1 & 0 & 0 \\ 0 & e^{i\phi_1} & 0 \\ 0 & 0 & e^{i\phi_2} \end{pmatrix},
\end{equation}
with two ``Majorana-like'' phases $\phi_1$, $\phi_2$ and the ``Dirac-like'' matrix
\begin{equation}
V_\text{PMNS} = \begin{pmatrix}  1 & 0 & 0 \\ 0 & c_{23} & s_{23} \\ 0 & -s_{23} & c_{23} \end{pmatrix} \begin{pmatrix} c_{13} & 0 & s_{13} e^{-i\delta} \\ 0 & 1 & 0 \\ -s_{13} e^{i\delta} & 0 & c_{13} \end{pmatrix} \begin{pmatrix} c_{12} & s_{12} & 0 \\ -s_{12} & c_{12} & 0 \\ 0 & 0 & 1 \end{pmatrix}.
\label{eq:definitionmatrixV}
\end{equation}
We have used here the abbreviations $c_{ij} = \cos(\theta_{ij})$, $s_{ij} = \sin(\theta_{ij})$ with mixing angles $\theta_{12}$, $\theta_{13}$, $\theta_{23}$, and the CP-violating phase $\delta$. 

It is clear that the ``Majorana-like'' phases $\phi_1$ and $\phi_2$ do not enter the projection operators \eqref{eq:projectionOperatorsLeptonMass}. In fact, one can write (no summation over the index $j$)
\begin{equation}
P_j = | \tilde l_j^+ \rangle  \langle \tilde l_j^+ | = (\Phi^\dagger_\text{PMNS})_{jk} (V_\text{PMNS}^\dagger)_{k\alpha} | l_\alpha^+ \rangle \langle l^+_\gamma | (V_\text{PMNS})_{\gamma m} (\Phi_\text{PMNS})_{mj} = (V_\text{PMNS}^\dagger)_{j\alpha} | l_\alpha^+ \rangle \langle l^+_\gamma | (V_\text{PMNS})_{\gamma j},
\label{eq:projectionoperatorsleptonmassbasis}
\end{equation}
and the matrix $\Phi_\text{PMNS}$ drops out completely. This shows that the phases $\phi_1$ and $\phi_2$ can not be constrained by the above method. However, the projectors $P_j$ depend on the remaining parameters $\delta$,  $\theta_{12}$, $\theta_{13}$ and $\theta_{23}$. 

In reality, we do not know the exact values of these four parameters, and we can make only an educated guess, 
$\hat \delta$, $\hat \theta_{12}$, $\hat \theta_{13}$, $\hat \theta_{23}$,
based on the available
constraints from neutrino oscillation and other experiments (where the three $\theta$ angles are much more constrained than the
CP-violating phase $\delta$). That is, we define an observable $Q$ in the lepton sector with possible outcomes $q=1,2,3$ such that the corresponding projection operators are defined analogous to \eqref{eq:projectionoperatorsleptonmassbasis},
\begin{equation}
\hat P_q = (V(\hat \delta, \hat \theta_{12}, \hat \theta_{13}, \hat \theta_{23})^\dagger)_{q\alpha} | l_\alpha^+ \rangle \langle l_\gamma^+ | (V(\hat \delta, \hat \theta_{12}, \hat \theta_{13}, \hat \theta_{23}))_{\gamma q}.
\label{eq:projectorPhat}
\end{equation}
Here, $V(\hat \delta, \hat \theta_{12}, \hat \theta_{13}, \hat \theta_{23})$ is our guess for the PMNS matrix, based on the chosen 
parameter values. 

We could now go on and calculate how the flavor oscillations of the neutrinos after a measurement of $Q$ on the lepton
(i.e. projection with 
$\hat P_q$) depend on the choice of parameters $\hat \delta$, $\hat \theta_{12}$, $\hat \theta_{13}$, $\hat \theta_{23}$ (we know that
the oscillations vanish if the correct values $\delta$,  $\theta_{12}$, $\theta_{13}$ and $\theta_{23}$ have been chosen).
This would enable us to constrain the PMNS matrix with this experiment. 

However, for simplicity and also in order to make some points regarding quantum information, we will first make the extra assumption that we are somehow able to measure neutrino masses directly. (In this case, the three $\theta$ angles could be also determined directly by simply correlating flavor and mass, i.e. by measuring $|U_{\alpha k}|^2$. The CP-violating phase $\delta$ however remains largely open.)

If one now tries to use the measurement result $q$ on the lepton to predict the outcome of a measurement of mass on the side of the neutrino, one must expect that the prediction fails in some cases, because the projection operator $\hat P_q$ is only an approximation
to the one based on the ``true" PMNS matrix.
It is useful to quantify this in terms of a conditional information entropy $H(\text{Neutrino }M|\text{Lepton } Q)$. Technically, this can be defined in terms of the conditional probability for the neutrino mass state $M=m$ given the result $Q=q$ from the measurement on the lepton, $p(m|q)$. More concrete, we define first the relevant information content gained from a measurement of $M$ given $Q=q$, as $i(m|q) = - \ln p(m|q)$. The corresponding information entropy is
\begin{equation}
H(\text{Neutrino } M|\text{Lepton }Q=q) = - \sum_m p(m|q) \ln p(m|q).
\end{equation}
The conditional entropy is defined as an expectation value of this quantity also with respect to different possibilities for $q$,
\begin{equation}
H(\text{Neutrino } M|\text{Lepton }Q) = \sum_q p(q) H(\text{Neutrino }M|\text{Lepton }Q=q) = - \sum_{m, q} p(m,q) \ln p(m| q),
\end{equation}
where we have used the joint probability $p(m,q)$ and the relation $p(m,q) = p(q) p(m|q)$ in the last equation. Note that by construction $H(\text{Neutrino } M|\text{Lepton } Q)\geq 0$ and $H(\text{Neutrino } M|\text{Lepton }Q)= 0$ precisely if and only if a measurement of $Q$ on the lepton leads to complete predictability of the mass measurement on the neutrino. 

Note that one can write
\begin{equation}
H(\text{Neutrino }M|\text{Lepton }Q) = - \sum_{m,j} p(m,j) \ln\left[ p(m,j) / p(j) \right] = H(\text{Neutrino }M, \text{Lepton }Q) - H(\text{Lepton } Q).
\label{eq:conditionalInformationEntropyDifference}
\end{equation}
In the last equation we use the joint entropy
\begin{equation}
H(\text{Neutrino }M, \text{Lepton }Q) = - \sum_{m,q} p(m,q) \ln p(m,q),
\end{equation}
and the reduced entropy for the quantity $Q$ based on the marginalized probabilities $p(q)=\sum_m p(m,q)$,
\begin{equation}
H(\text{Lepton }Q) = - \sum_q p(q) \ln p(q).
\end{equation}

For the fully entangled state \eqref{eq:fullyentangledstate} one has always $p(q)=1/3$ and therefore $H(\text{Lepton }Q) =\ln (3)$. Moreover, {\it if} the parameters defining $Q$ correspond to the real physical values, $\hat \delta=\delta$, $\hat \theta_{12} = \theta_{12}$, $\hat \theta_{13}= \theta_{13}$ and $\hat \theta_{23}=\theta_{23}$, one has the joint distribution $p(m,q)=\frac{1}{3}\delta_{mq}$ and thus $H(\text{Neutrino }M, \text{Lepton }Q) = \ln(3)$. In this case, the conditional entropy vanishes as expected, $H(\text{Neutrino }M|\text{Lepton }Q) =0$. 

More general, the joint probability is given for a state described by the density matrix $\rho$ by
\begin{equation}
p(m,q) = \text{Tr} \left\{ P^\text{Neutrino}_m \hat P^\text{Lepton}_q  \rho \right\}
\end{equation}
For the entangled pure state $\rho=|\psi \rangle \langle \psi |$ using \eqref{eq:fullyentangledstate} and the projector \eqref{eq:projectorPhat} one can calculate this explicitly and finds
\begin{equation}
p(m,q) = \frac{1}{3} \left|(V_\text{PMNS}^\dagger)_{m\alpha} V_{\alpha q}\right|^2.
\end{equation}
This is compatible with the statement above for $V=V_\text{PMNS}$.

\begin{figure}[t]
\centering
\includegraphics[width=0.44\textwidth]{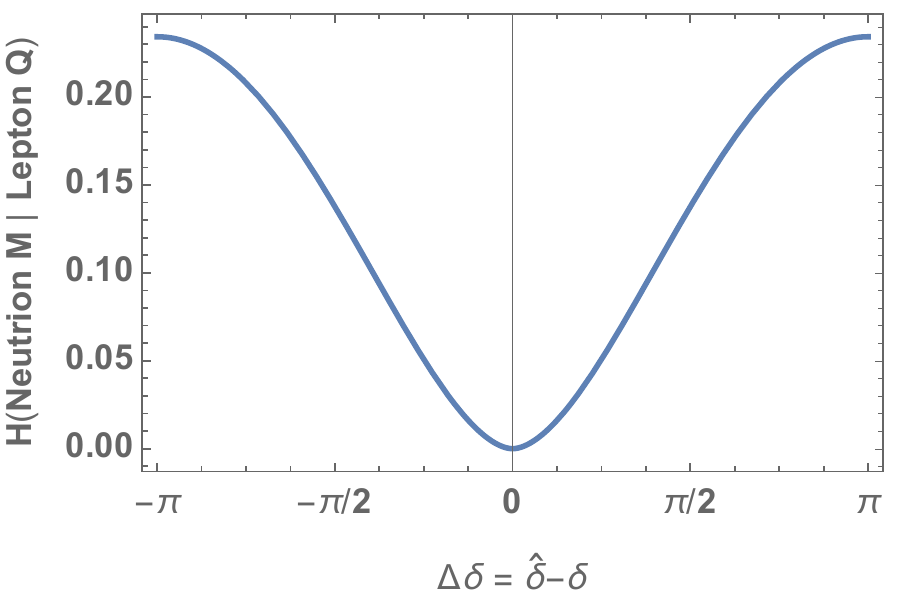}
\caption{Conditional information entropy $H(\text{Neutriono } M| \text{Lepton } Q)$ for a measurement of neutrino mass $M$ after a projective measurement $Q$ on a lepton in a state that is fully entangled with the neutrino. The operator $Q$ defined in eq.\ \eqref{eq:projectorPhat} depends on the phase parameter $\hat \delta$, or more precisely its difference to the physical CP violating phase $\Delta\delta=\hat \delta - \delta$. The conditional entanglement entropy assumes its minimal value $H(\text{Neutriono } M| \text{Lepton } Q)=0$ for $\Delta \delta=0 \text{ mod }2\pi$. Finding the global minimum of this conditional information entropy allows to find the physial CP violating phase $\delta$ and more general the physical matrix $V_\text{PMNS}$.}
\label{fig1}
\end{figure}

Let us illustrate this by assuming that the current best fit parameters $\theta_{12} = 33.62^\circ$, $\theta_{13}=8.54^\circ$, $\theta_{23}=47.2^\circ$ and $\delta=234^\circ$ \cite{Esteban:2016qun} correspond in fact to the real physical values. We can then determine $H(\text{Neutriono } M| \text{Lepton } Q)$ as a function of $\Delta\delta=\hat \delta - \delta$. The result is shown in fig.\ \ref{fig1}. 
One observes that indeed $H(\text{Neutriono } M| \text{Lepton } Q)=0$ for $\Delta\delta=0 \text{ mod }2\pi$ and that $H(\text{Neutriono } M| \text{Lepton } Q)>0$ otherwise. A similar picture holds in fact if also the angles $\hat \theta_{12}$, $\hat \theta_{13}$ and $\hat \theta_{23}$ are deviating from their (assumed) physical values. In this sense one could use the conditional information entropy $H(\text{Neutriono } M| \text{Lepton } Q)$ to find the physical configuration for $V_\text{PMNS}$. One needs to try different values $\hat \theta_{12}$, $\hat \theta_{13}$, $\hat \theta_{23}$ and $\hat \delta$, and the physical values are obtained for $H(\text{Neutriono } M| \text{Lepton } Q)=0$.

Of course, in practice it is rather difficult to determine the mass of a neutrino directly. It might be much simpler to detect the absence of neutrino oscillations. It is clear that a state with definite neutrino mass such that $H(\text{Neutriono } M| \text{Lepton } Q)=0$ has no neutrino oscillations. What is less clear is the opposite statement: Does the absence of neutrino oscillations imply a state of definite mass?

We will address this question in the specific context of the entanglement-based experiment described above and more specifically consider again the CP violating phase. Let us assume for simplicity that the mixing angles $\theta_{12}$, $\theta_{13}$ and $\theta_{23}$ are known and let us fix the projection operator \eqref{eq:projectorPhat} for the observable $Q$ such that $\hat\theta_{12}=\theta_{12}$, $\hat\theta_{13}=\theta_{13}$ and $\hat\theta_{23}=\theta_{23}$. It remains to vary the CP phase parameter $\hat\delta$ or, equivalently, the difference to the physical value $\Delta\delta=\hat\delta-\delta$. The question is then whether the presence or absence of neutrino flavor oscillations is sensitive to $\Delta\delta$. 

To formalize this question, we consider the conditional probability $p_{\Delta t}(\alpha|q)$ to find the neutrino in the flavor state $\alpha\in\{e,\mu,\tau\}$ after some time $\Delta t$ (after the initial preparation of the entangled state \eqref{eq:leptonneutrinoentangledstate}) given that the projective measurement of the observables $Q$ on the lepton led to the result $q$. One can write $p_{\Delta t}(\alpha|q) = p_{\Delta t}(\alpha, q) / p(q)$ with the marginalized probability $p(q)=1/3$ (independent of time) and the joint probability
\begin{equation}
p_{\Delta t}(\alpha,q) = \text{tr} \left\{ P^\text{Neutrino}_\alpha(\Delta t) \hat{P}^\text{Lepton}_q \rho \right\}.
\end{equation}
Using 
\begin{equation}
P^\text{Neutrino}_\alpha = | \nu_\alpha(\Delta t) \rangle \langle \nu_\alpha(\Delta t) | = \sum_{j,k,\beta,\lambda} U^\dagger_{j\alpha}  \, e^{-iE_j \Delta t} \, U_{\beta j} \, | \nu_\beta(0) \rangle \langle \nu_\lambda(0) | \, U^\dagger_{k\lambda} \, e^{iE_k \Delta t} \, U_{\alpha k} ,
\end{equation}
the operator $\hat P_q$ from eq.\ \eqref{eq:projectorPhat} and the entangled state \eqref{eq:leptonneutrinoentangledstate}, one finds after some straight forward algebra,
\begin{equation}
p_{\Delta t}(\alpha| q) = \sum_{j,k}  e^{-i(E_j-E_k)\Delta t} \; c_{\alpha q,jk},
\label{eq:condprobexpansion}
\end{equation}
where
\begin{equation}
c_{\alpha q,jk} = \sum_{\beta,\lambda} V^\dagger_{q\beta} (V_\text{PMNS})_{\beta j} (V_\text{PMNS})^\dagger_{j\alpha} (V_\text{PMNS})_{\alpha k} (V_\text{PMNS})^\dagger_{k\lambda} V_{\lambda q},
\end{equation}
and in the last equation we have used the abbreviation $V_{\lambda q} = V_{\lambda q}(\hat \delta, \hat \theta_{12}, \hat \theta_{13}, \hat \theta_{23})$.
Note first that $c_{\alpha q,jk}=c^*_{\alpha q,kj}$ such that $p_{\Delta t}(\alpha| q)\in \mathbbm{R}$ as it should be. Also, for $V=V_\text{PMNS}$ one has $c_{\alpha q,jk}=\delta_{qj} \delta_{qk} |(V_\text{PMNS})_{\alpha q}|^2$ such that $p_{\Delta t}(\alpha| q) = | (V_\text{PMNS})_{\alpha q}|^2$ is independent of time $\Delta t$, i.\ e.\ there are no flavor oscillations. In a similar way, one finds in general for $\Delta t=0$ the result $p_{\Delta t=0}(\alpha| q) = | V_{\alpha q}|^2$. Flavor oscillations are present whenever one (or several) of the off-diagonal coefficients $c_{\alpha q,12}$, $c_{\alpha q,13}$ and $c_{\alpha q,23}$ are non-vanishing.  
\begin{figure}[t]
\centering
\includegraphics[width=0.44\textwidth]{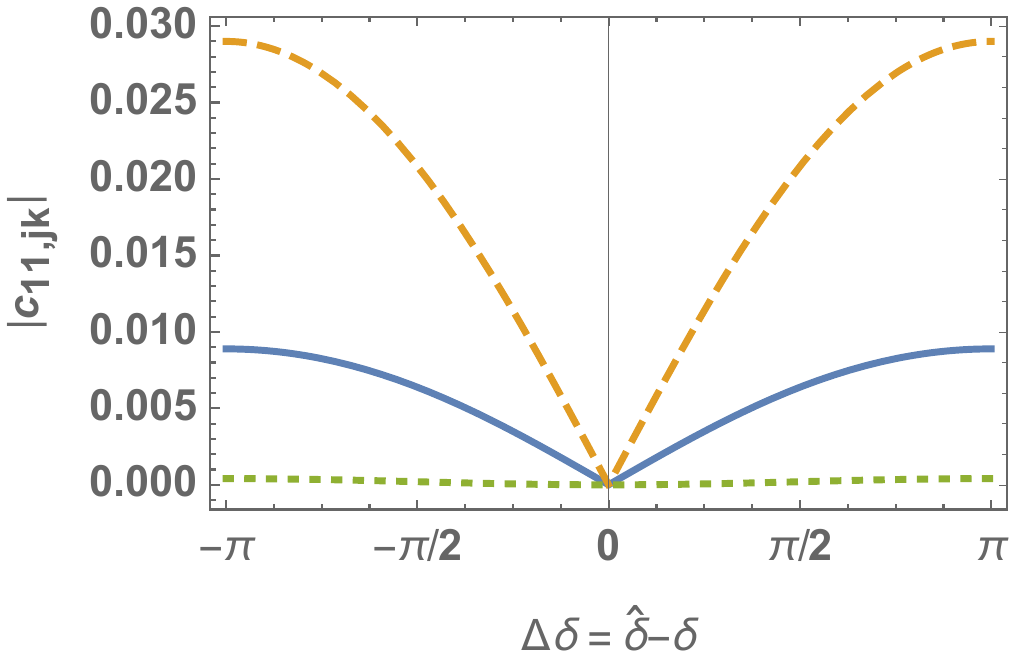}
\caption{Neutrino flavor oscillation amplitudes $|c_{11,12}|$ (solid line), $|c_{11,13}|$ (dashed line) and $|c_{11,23}|$ (dotted line) in the conditional probability $p_{\Delta t}(\alpha=1 | q=1)$ to find the neutrino flavor $\alpha=1$ provided the observable $Q$ on an entangled lepton gave the outcome $q=1$. The expansion coefficients $c_{\alpha q,jk}$ are defined in eq.\ \eqref{eq:condprobexpansion}. The result depends on the difference between the angle $\hat \delta$ that enters the operator $Q$ according to eq.\ \eqref{eq:projectorPhat} and the physical CP violating angle $\Delta\delta = \hat\delta - \delta$. For $\Delta\delta=0$ all flavor oscillations disappear.}
\label{fig2}
\end{figure}

As an example, we consider the case $\alpha=1$, $q=1$ and show in fig.\ \ref{fig2} the absolute values of the expansion coefficients $|c_{11,12}|$ (solid line), $|c_{11,13}|$ (dashed line) and $|c_{11,23}|$ (dotted line) as a function of the difference between angles $\Delta\delta = \hat \delta - \delta$. One observes that the conditional probability $p_{\Delta t}(\alpha=1,q=1)$ clearly shows oscillations, except if the parameter $\hat\delta$ entering the operator $Q$ for the projective measurement on the lepton agrees with the real physical CP violating phase angle $\delta$ such that $\Delta\delta=0$. We have checked that all flavor oscillations disappear for $\Delta\delta=0$ and are mostly non-vanishing otherwise. This underlines that the absence of neutrino oscillations after a projective measurement on the lepton in an overall entangled state can be used to measure the CP violating phase and actually the entire PMNS matrix.

Let us stress that it is important for the above protocol to consider the conditional probability $p_{\Delta t}(\alpha|q)$ or the joint probability $p_{\Delta t}(\alpha, q)$. The marginalized probability $p_{\Delta t}(\alpha)$ to find a neutrino in a flavor state $\alpha$, marginalized over the result of any measurement on the lepton, is actually $p_{\Delta t}(\alpha)=1/3$, independent of $\Delta t$ and flavor species $\alpha$. This follows from the reduced density matrix for the neutrino in the state \eqref{eq:leptonneutrinoentangledstate},
\begin{equation}
\rho_\text{Neutrino} = \sum_\alpha \langle l^+_\alpha |\psi\rangle \langle \psi | l^+_\alpha \rangle =  \frac{1}{3} \sum_\beta | \nu_\beta \rangle \langle \nu_\beta |.
\end{equation} 
Because it is proportional to the unit matrix, this density matrix is actually invariant under unitary transformations and therefore keeps its form in any basis and also as a function of time.

\subsection{Entropic uncertainty relation with quantum memory}
\label{sec:Entropicuncertaintyrelationwithquantummemory}

To formulate uncertainty relations with quantum memory, one needs the concept of {\it quantum conditional entropies} \cite{Coles:2017,Wilde}. First, the classical-quantum conditional entropy for a variable $\hat z$ measured on the subsystem $A$ with a complete basis $|z_j\rangle$ with side information from a measurement from the entangled quantum system $B$ is defined in analogy to \eqref{eq:conditionalInformationEntropyDifference} by
\begin{equation}
H(z|B) =  - \sum_j \text{Tr}_B \left\{ \langle z_j | \rho_{AB} | z_j\rangle \ln \langle z_j | \rho_{AB} | z_j \rangle \right\} - S(\rho_B).
\label{eq:defclassicalquantum}
\end{equation}
Here $\rho_{AB}$ is the density matrix for the full system $A+B$ and $\rho_B=\text{tr}_A\{ \rho_{AB} \}$ is the reduced density matrix for subsystem $B$ with von Neumann entropy $S(\rho_B)$. Note that for product states $\rho_{AB} = \rho_A \rho_B$, one has $H(z|B)=H^{(z)}$. One can show that $H(z|B) \geq 0$ (see exercise 11.9.4 in ref.\ \cite{Wilde}).
In a similar way one can define the classical-quantum conditional entropy for the measurement of $\hat x$ on subsystem $A$ with quantum side information from $B$ denoted by $H(x|B)$. 

In addition we introduce the {\it quantum conditional entropy} as a difference of von Neumann entropies
\begin{equation}
S(A|B) = S(\rho_{AB}) - S(\rho_B).
\end{equation}
Note that this quantity can be negative as a result of entanglement, for example if the full system is in a pure state with $S(\rho_{AB})=0$.

The extension of the entropic uncertainty relation \eqref{eq:FrankLieb2002} to situations with quantum memory reads then \cite{Berta}
\begin{equation}
H(z|B) + H(x|B) \geq 2 \ln (1/c) + S(A|B).
\label{eq:EURquantumsideinformation}
\end{equation}

If the system $A$ is sufficiently entangled with the system $B$ carrying the quantum side information, and if the total density matrix $\rho_{AB}$ is pure enough, one has $S(A|B) \leq 0$ and the right hand side of \eqref{eq:EURquantumsideinformation} is smaller than in a situation without side information. This demonstrates quantitatively how quantum side information in the form of entanglement can be used to reduce measurement uncertainties.

\section{Conclusions}
For finite-dimensional Hilbert spaces, like the three-dimensional mass/flavor space $\mathcal{H}_f$ of neutrinos,
the information-entropic version of uncertainty is, in many cases, better suited than the Heisenberg uncertainties
that are defined in terms of standard deviations. We have demonstrated the difference between these two concepts
and discussed the associated uncertainty relations.

We suggest that information-theoretic concepts have interesting applications in the context of neutrino physics.
In particular, we identified a number of areas where they may be applied. Quantum computation and communication, 
the standard arena for these concepts, can, at least in principle, be realized with $\mathcal{H}_f$, using
entangled lepton/neutrino pairs. Entropies associated with neutrino masses may play a role in cosmology:
for hypothetical massive right-handed neutrinos in the very early universe, and for the left-handed neutrinos in
the present era.

We also discussed phenomena related to quantum entanglement in the neutrino / lepton sector of the standard model. In particular, entangled pairs of neutrinos and leptons arise naturally as a result of electroweak decays. One can then use ``side information'' from a projective measurement of the lepton to predict results of subsequent measurements on the neutrino. The amount of knowledge gained from such a ``quantum memory'' can be quantified in terms of conditional information entropies. As a particularly interesting application of this set of ideas we have discussed a possibility to determine the CP-violating phase of the PMNS matrix. To that end one performs a projective measurement consisting of a unitary evolution operator that depends on phase $\hat \delta$ and a measurement on a lepton in a state that is fully entangled with a neutrino. If the parameter $\hat\delta$ agrees with the physical CP violating phase $\delta$ of the PMNS matrix, 
$\hat\delta=\delta$, the neutrino is predicted to be in a mass eigenstate. Experimentally, this can be verified by the absence of neutrino oscillations. More generally, this method can be applied also to determine the non CP-violating angles $\theta_{12}$, $\theta_{13}$ and $\theta_{23}$ and thereby the entire ``Dirac-like'' part of the PMNS matrix. 

We finally also discussed briefly an extension of the entropic uncertainty relations to situations with quantum memory. In summary, we believe that quantum information theoretic concepts and notions, in particular entropic uncertainty relations and quantum entanglement, can be applied fruitfully in neutrino physics.

\section*{Acknowledgment}
The work of S.\ F.\ is part of and supported by the DFG Collaborative Research Centre ``SFB 1225 (ISOQUANT)''.

\end{document}